\documentclass[%
 reprint,superscriptaddress,
 amsmath,amssymb,prx]{revtex4-1}

\usepackage{physics}
\usepackage{xcolor}
\usepackage{xtab,afterpage,longtable}
\usepackage{lipsum}
\usepackage{graphicx}
\usepackage{dcolumn}
\usepackage{booktabs} 
\usepackage{multirow}
\usepackage{color}
\usepackage{hyperref}
	
\newcommand{\RomanNumeralCaps}[1]
    {\MakeUppercase{\romannumeral #1}}
    
\begin{document}

\title{Reinforcement learning-guided long-timescale simulation of hydrogen transport in metals}

\author{Hao Tang}
\affiliation{Department of Materials Science and Engineering, Massachusetts Institute of Technology, MA 02139, USA}

\author{Boning Li}
\affiliation{
   Research Laboratory of Electronics, Massachusetts Institute of Technology, Cambridge, MA 02139, USA}
\affiliation{Department of Physics, Massachusetts Institute of Technology, MA 02139, USA}

\author{Yixuan Song}%
\affiliation{Department of Materials Science and Engineering, Massachusetts Institute of Technology, MA 02139, USA}

\author{Mengren Liu}%
\affiliation{Department of Materials Science and Engineering, Massachusetts Institute of Technology, MA 02139, USA}

\author{Haowei Xu}%
\affiliation{
   Department of Nuclear Science and Engineering, Massachusetts Institute of Technology, Cambridge, MA 02139, USA}

\author{Guoqing Wang}%
\affiliation{
   Research Laboratory of Electronics, Massachusetts Institute of Technology, Cambridge, MA 02139, USA}
\affiliation{
   Department of Nuclear Science and Engineering, Massachusetts Institute of Technology, Cambridge, MA 02139, USA}

\author{Heejung Chung}%
\affiliation{Department of Materials Science and Engineering, Massachusetts Institute of Technology, MA 02139, USA}

\author{Ju Li}%
 \email{liju@mit.edu}
 \affiliation{Department of Materials Science and Engineering, Massachusetts Institute of Technology, MA 02139, USA}
\affiliation{
   Department of Nuclear Science and Engineering, Massachusetts Institute of Technology, Cambridge, MA 02139, USA}
   
\date{\today}
             
\begin{abstract}
Atomic diffusion in solids is an important process in various phenomena. However, atomistic simulations of diffusion processes are confronted with the timescale problem: the accessible simulation time is usually far shorter than that of experimental interests. In this work, we developed a long-timescale method using reinforcement learning that simulates diffusion processes.  
As a testbed, we simulate hydrogen diffusion in pure metals and a medium entropy alloy, CrCoNi, getting hydrogen diffusivity reasonably consistent with previous experiments. We also demonstrate that our method can accelerate the sampling of low-energy configurations compared to the Metropolis-Hastings algorithm using hydrogen migration to copper (111) surface sites as an example.
\end{abstract}

\maketitle
\section{Introduction}
Diffusional atomic motion is an essential microscopic process in the kinetics of materials~\cite{allen2005kinetics}. Various interesting phenomena and applications are rooted in diffusion-related processes, from the interdiffusion at metal interfaces, vacancy and void formation, to hydrogen embrittlement~\cite{dwivedi2018hydrogen} and resistance switching in oxide memristors~\cite{kumar2017oxygen}. One important tool to study the diffusion process is atomistic simulation~\cite{uberuaga2004structure, PhysRevLett.65.729}, which can simulate a wide range of materials phenomena~\cite{li2002atomistic,li2011diffusive}. However, a critical challenge of atomistic simulation of diffusion-related process is the timescale problem~\cite{uberuaga2020computational}: the atomic vibration has a timescale of fs - ps; however, the diffusion-related transitions between adjacent energy minima have orders of magnitude larger timescale. That is because the energy barriers on the diffusion pathway slow down the diffusion process~\cite{li2011diffusive}. The timescale problem limits most of the straightforward molecular dynamics simulations to nanoseconds, which fall short of the timescales relevant to many diffusion-related phenomena~\cite{uberuaga2020computational,perez2018long}. Therefore, different methods are needed to deal with the long-timescale problem~\cite{uberuaga2020computational}.

Our work is based on one of the widely studied algorithms, the kinetic Monte Carlo (KMC) method~\cite{voter2007introduction}, where one directly works with diffusion timescale without explicitly showing the vibration timescale motion. Traditional KMC (in contrast with off-lattice KMC) requires energy minima and transition pathways (the so-called event table) as input.  
However, as the diffusion pathway is sometimes counter-intuitive, correctly determining the necessary input information of KMC is not a trivial task~\cite{voter2007introduction}. 
To conduct a simulation without a known event table, the off-lattice KMC is developed~\cite{trochet2020off}. The algorithm conducts saddle point searches to obtain the diffusion pathways along with the KMC simulation. Another method reported to have advantageous efficiency is temperature accelerated dynamics (TAD), where the transition pathways are explored by high-temperature molecular dynamics~\cite{voter2000temperature}. In both methods, the transition pathway is explored by random sampling (random initial guess in the saddle point search for off-lattice KMC, and random thermal motion for TAD). However, as the configuration space is high-dimensional, it requires a large amount of random sampling to be confident that the correct transition pathway is obtained, which limits the simulation system size and accessible timescale~\cite{trochet2020off}.

In this work, we developed a reinforcement learning (RL) method that guides the transition pathway sampling in off-lattice KMC to simulate long-timescale diffusion processes. Instead of searching for all nearby saddle points along randomly sampled initial directions~\cite{trochet2020off}, we use parameterized neural network model to guide the saddle-point search. The model can predict the direction of atomic motion that yields the high-probability transition pathway. That avoids the repeated saddle-point searches, which is the most significant contributor to the computational cost of the off-lattice KMC. We demonstrate that our RL model can either simulate physical diffusion trajectories or sample low-energy configurations in complex energy landscapes by simulating the hydrogen diffusion in alloys and metal surfaces. 
\section{Results}
\begin{figure}[t]
\centering
\includegraphics[width=\linewidth]{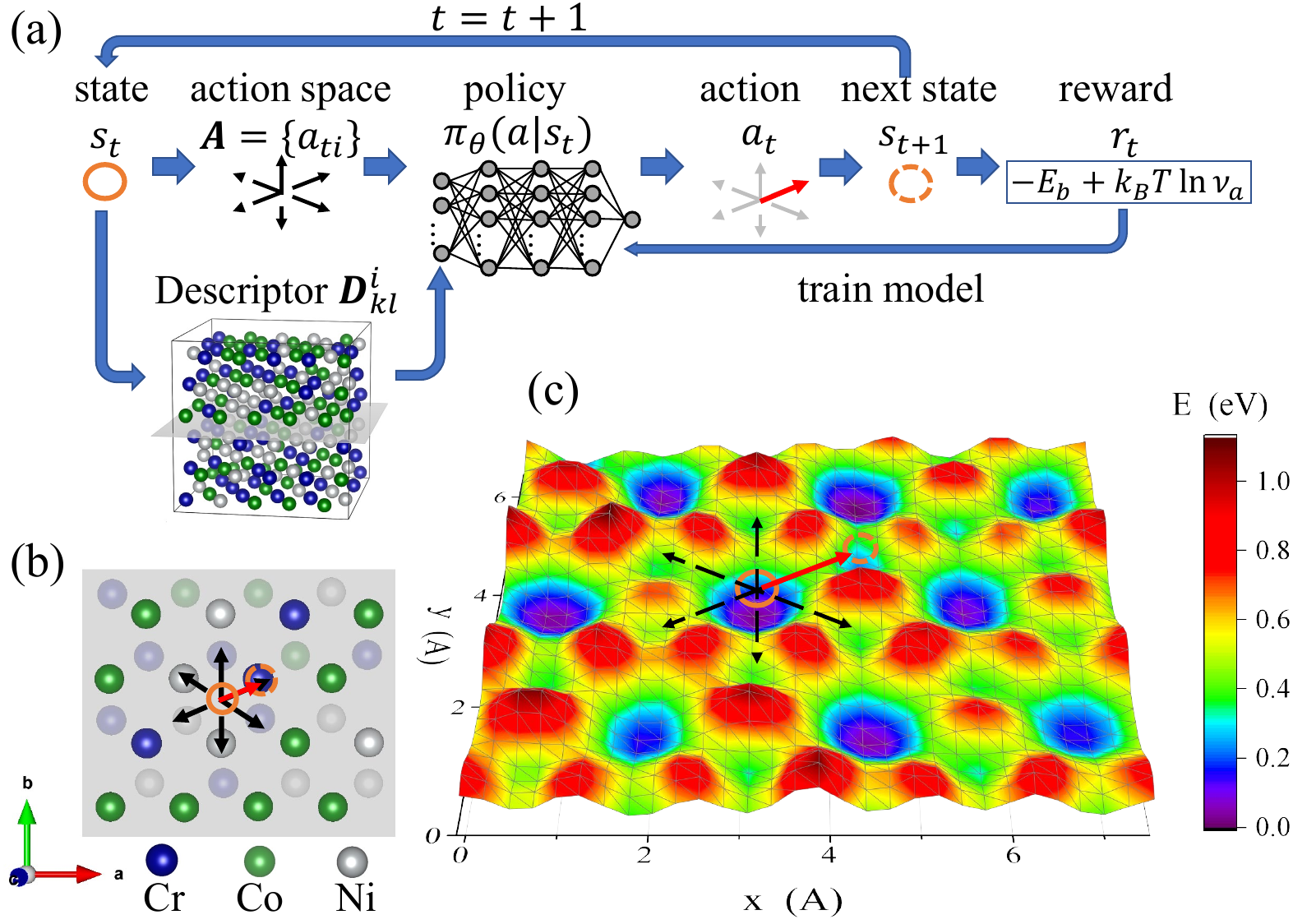}
\caption{(a) Computational workflow of the RL long timescale method illustrated on (b) hydrogen diffusion in CrCoNi medium entropy alloy. The blue, green, and grey spheres represent Cr, Co, and Ni atoms, respectively. The orange circle, black dashed arrow, and red arrow represent state, action space, and selected action, respectively. (c) The potential energy landscape of a hydrogen atom on the grey planes in (a, b). When calculating the energy, surrounding atoms and the $z$-coordinate of the hydrogen atoms are relaxed.}
\label{fig:intro}
\end{figure}

Here, we briefly describe our RL method, as illustrated in Fig.~\ref{fig:intro}a. In atomic diffusion, the energy landscape has a large number of local minima separated by transition energy barriers. In this paper, we use hydrogen diffusion in face-centered cubic (FCC) alloys as an example, as shown in Fig.~\ref{fig:intro}b. In the local energy minimum configurations of FCC bulk structures, hydrogen atoms reside in octahedral and tetrahedral interstitial sites shown as the deep blue and shallow green potential wells in Fig.~\ref{fig:intro}c, where the octahedral site has lower energy. The energy landscape is provided by a universal neural network PreFerred Potential (PFP)~\cite{TakamotoOLL23} throughout this paper. Beginning from a given local energy minimum configuration or ``state" $s_t = (\vec{r}_1,\vec{r}_2,\cdots ,\vec{r}_N)$ (the orange circles in Fig.~\ref{fig:intro}, where $\vec{r}_i$ is the coordinates of the $i$th atom), a set of possible transition displacements $\{a_{ti}\}$ (also called ``actions") are first identified. In our problem, this is realized by identifying the polyhedron surrounding each hydrogen atom formed by its metal neighbors where possible actions are defined by translations through all face centers of the polyhedron (See section \RomanNumeralCaps{4}.A for details). 

In the next step, an action $a_t$ is selected from the action space $\mathcal A_{s_t}\equiv \{a_{ti}\}$ based on the atomic descriptor $\mathcal D$ of the configuration $s_t$. The probability of selecting each action $a$, $\pi_\theta (a|s_t)$, is given by the Boltzmann policy based on a neural-network value function $Q_\theta (s,a)$~\cite{wang2013optimization}:
\begin{equation}
    \pi_\theta (a|s_t) = \frac{e^{Q_\theta (s_t,a)/k_{\rm B}T}}{\sum_{a'\in \mathcal A_{s_t}}e^{Q_\theta (s_t,a')/k_{\rm B}T}},
    \label{eq:policy}
\end{equation}
where $\theta$ represents the model parameters, $k_{\rm B}$ and $T$ are the Boltzmann constant and temperature. $Q_\theta (s,a)$ can also depend on $T$ if the vibrational entropy contribution is considered, which will be discussed later. 

After selecting an action $a_t = (i,\vec{v})$, the $i$th atom is displaced by vector $\vec{v}$ across the energy barrier. The system is then relaxed to the next state, $s_{t+1}$, using the MDMin algorithm implemented in the Atomistic Simulation Environment~\cite{larsen2017atomic}. Parameters of the transition, including the transition energy barrier $E_b^{\text{\rm NEB}}$, the attempt frequency $\nu_a^{\text{\rm NEB}}$, and the energy change after the transition $\Delta E$, can then be estimated using the NEB method~\cite{jonsson1998nudged} setting $s_t$ and $s_{t+1}$ as the initial and final points. The reward of this transition, $r_t$, is designed to encourage either reproducing transition probabilities of the harmonic transition state theory (HTST)~\cite{Asgeirsson2020} or an energy minimization strategy, which will be discussed in the next part. The whole simulation trajectory is produced by repeating the above scheme that generates the next state according to the current state.

The $Q_\theta (s, a)$ model is constructed based on the DeepPot-SE sub-networks~\cite{zhang2018end}. As the atomic interaction in alloys is short-range, we assume $Q_\theta (s, a=(i,\vec{v}))$ is a function of the atomic environment of the moved atom $i$ and its displacement $\vec{v}$. The descriptor $\mathcal D^i$ should be equivariant under translation, rotation, and permutation symmetry operations of the atomic system, realized by the following construction:
\begin{equation}
   \tilde{R}^i = 
   \left[
    \begin{matrix}
    \hat{\vec{r}}_{i1}\cdot \hat{\vec{r}}_{i1} & \cdots & \hat{\vec{r}}_{i1}\cdot \hat{\vec{r}}_{iM} & \hat{\vec{r}}_{i1}\cdot \vec{v}\\ 
    \hat{\vec{r}}_{i2}\cdot \hat{\vec{r}}_{i1} & \cdots & \hat{\vec{r}}_{i2}\cdot \hat{\vec{r}}_{iM} & \hat{\vec{r}}_{i2}\cdot \vec{v}  \\
    \vdots & \vdots &  &\vdots \\
    \hat{\vec{r}}_{iM}\cdot \hat{\vec{r}}_{i1} & \cdots & \hat{\vec{r}}_{iM}\cdot \hat{\vec{r}}_{iM} & \hat{\vec{r}}_{iM}\cdot \vec{v} \\
    \vec{v}\cdot \hat{\vec{r}}_{i1} & \cdots & \vec{v}\cdot \hat{\vec{r}}_{iM}& |\vec{v}|^2\\ 
    \end{matrix}
    \right],
\end{equation}
\begin{equation}
    \mathcal D^i_{kl} = \sum_{m,n=1}^{M+1} G^1_k(f_{c}(r_{im}),c_m)\tilde{R}^i_{mn}G^2_l(f_{c}(r_{in}),c_n),
    \label{eq:descriptor}
\end{equation}
where $\hat{\vec{r}}_{ij}\equiv \frac{f_c(r_{ij})\vec{r}_{ij}}{r_{ij}}, \vec{r}_{ij} \equiv \vec{r}_j - \vec{r}_i$, $r_{ij}\equiv |\vec{r}_{ij}|$, $j=1,2,\cdots ,M$ goes through all atoms around the $i$th atom within a cut-off radius $r_c$. $f_c(r)$ is a cutoff function as defined in Ref.~\cite{zhang2018end}, which goes smoothly to zero at a cutoff radius $r_c$, and $G_k^1$ and $G_l^2$ are embedding neural networks parametrized by $\theta_{\rm emb}$. $c_m (m = 1,2,\cdots, M)$ are the atomic species of the $m$th atom, and we set $c_{M+1}$ as a unique ``action species". The descriptor $\mathcal D^i$ is invariant under all symmetry operations. The descriptor is then flattened to a vector and passed to a multilayer perceptron (MLP) that outputs the $Q$ function: $Q_\theta (s, a=(i,\vec{v})) = {\rm MLP}_{\theta_{\rm fit}}(\mathcal D^i(\theta_{\rm emb}))$, where the model parameters $\theta = (\theta_{\rm fit}, \theta_{\rm emb})$ includes both parameters of the MLP $\theta_{\rm fit}$ and that of the embedding network $\theta_{\rm emb}$ (see section \RomanNumeralCaps{4}.B for detailed parameter settings).

\begin{table}[]
\renewcommand\arraystretch{1.4}
\caption{Hydrogen self-diffusion simulation results in pure copper, pure nickel, and CrCoNi medium entropy alloy. $\Delta E_b\equiv \sqrt{\langle (E_b^{\rm NN}-E_b^{\rm NEB})^2\rangle}$ and $\Delta \nu_a\equiv \sqrt{\langle (\ln{\nu_a^{\rm NN}}-\ln{\nu_a^{\rm NEB}})^2\rangle}$ are the validation error of model prediction on transition energy barrier and attempt frequency. The activation energy $Q$ and coefficient $D_0$ are fitted by reinforcement-learning-simulated diffusivity $D=D_0e^{-Q/k_{\rm B}T}$ using maximal-likelihood estimation, and $D_0^{\text{exp}}$ and $Q^{\text{exp}}$ are the values from previous experiments.}
\centering
\begin{tabular}{ p{2.6 cm}p{1.8cm}p{2cm}p{1.5cm}  }
    \hline
    \hline
         &  Cu & Ni & CrCoNi\\
    \hline
    \hline
    $\Delta E_b$ ({\rm eV}) & 0.020 & 0.022 & 0.037\\
    \hline
    $\Delta \ln{\nu_a}$ & 0.09 & 0.12 & 0.12\\
    \hline
    $D_0 (10^{-7} {\rm m}^2/{\rm s})$ & 3.6&  3.1 & 5\\
    \hline
    $Q ({\rm eV})$ & 0.30 &  0.33& 0.43\\
    \hline
    $D_0^{\text{exp}} (10^{-7} {\rm m}^2/{\rm s})$ & 3.69~\cite{sakamoto1982electrochemical}&  $ 0.15-6.98$~\cite{ansari2000determination}& --\\
    & 21.1~\cite{ISHIKAWA1985445} &  $ 1.1-6.87$~\cite{LEE1984859}& --\\
    & 17.4~\cite{magnusson2017diffusion}& &\\
    \hline
    $Q^{\text{exp}} ({\rm eV})$ & 0.38~\cite{sakamoto1982electrochemical} &  0.31-0.44~\cite{ansari2000determination}& --\\
     & 0.46~\cite{ISHIKAWA1985445} &  0.37-0.44~\cite{LEE1984859}& --\\
     & 0.435~\cite{magnusson2017diffusion} &&\\
    \hline
    \hline
    \end{tabular}
    \label{table}
\end{table}

\begin{figure}[t]
\centering
\includegraphics[width=\linewidth]{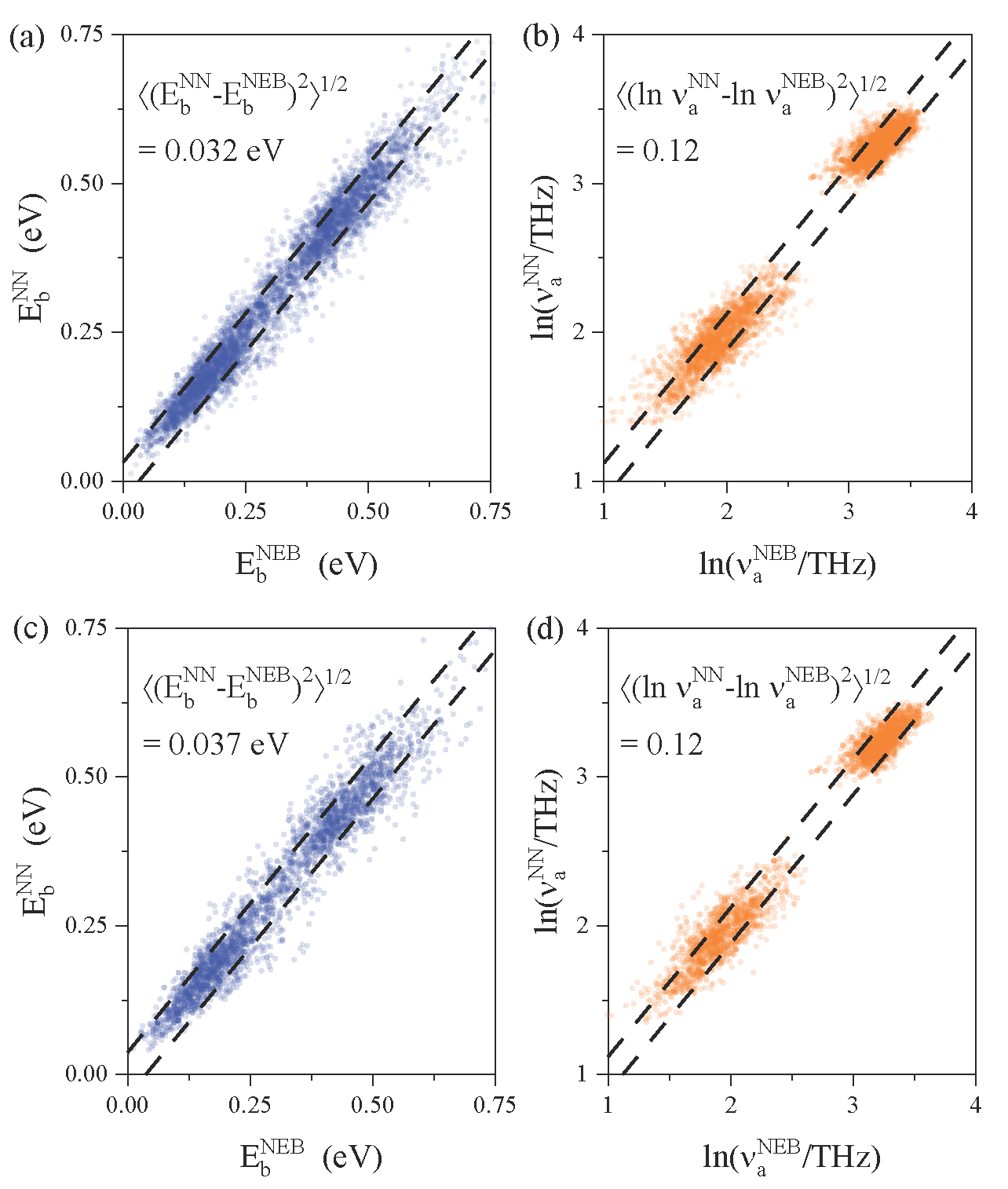}
\caption{Comparison of the neural network model prediction of (a) transition energy barriers $E_b^{\rm NN}$ and (b) attempt frequency $\nu_a^{\rm NN}$ with those calculated by the NEB method in the training dataset, $E_b^{\rm NEB}$ and $\nu_a^{\rm NEB}$. Validation on the testing dataset is shown in (c) and (d).}
\label{fig:validation}
\end{figure}

By choosing different reward functions, our method has two working modes: transition kinetics simulator (TKS) and low-energy states sampler (LSS). The TKS aims to simulate physical transition probabilities according to HTST, and the LSS aims to converge to global energy minimum configurations.

\begin{figure*}[t]
\centering
\includegraphics[width=\linewidth]{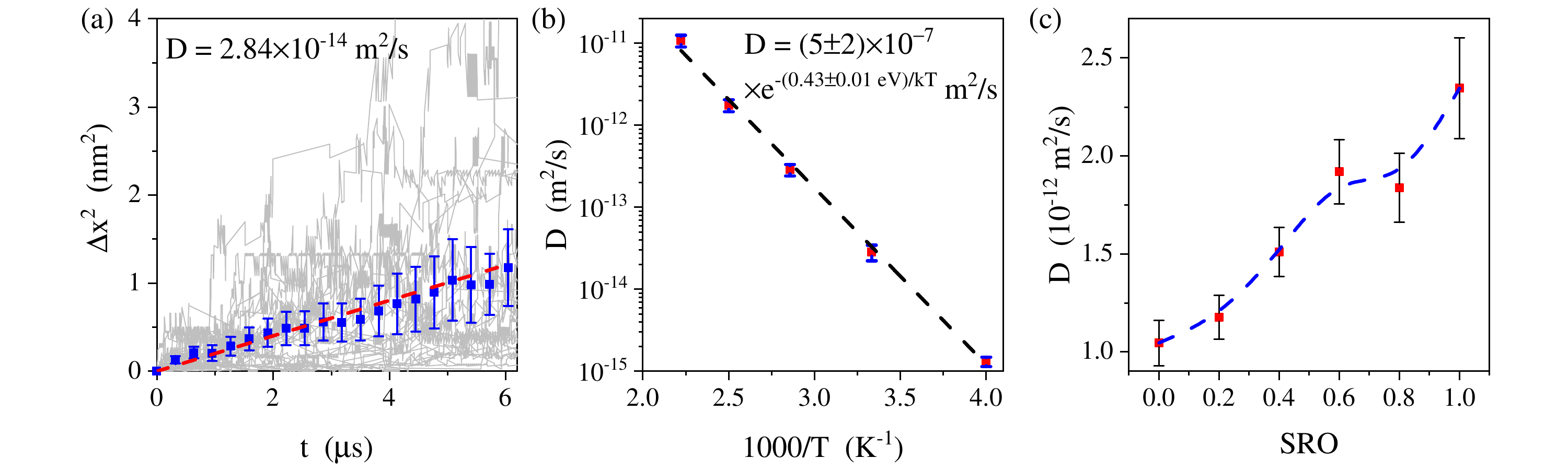}
\caption{Hydrogen diffusion simulation in CrCoNi medium entropy alloy. (a) Square hydrogen diffusion displacement $\Delta x^2$ (absolute value) as a function of time under 300 K. The grey lines show 30 trajectories; the blue squares and error bars are the mean square displacements and their error range ($\pm$ one standard error); the red dashed lines is the linear fitting of the blue dots. (b) Arrhenius plot of hydrogen self-diffusivity under different temperatures. The blue caps show the error bar of calculated diffusivities, and the black dashed line is a linear fitting of $\log{D}$ $vs$ $\frac{1000}{T}$. (c) Hydrogen self-diffusivity at 400 K as a function of the short-range ordering parameter (the dashed line is a B-spline~\cite{gordon1974b} connecting the data points). SRO = 0 corresponds to a fully random solid solution, SRO = 1 corresponds to WC parameters obtained from Ref.~\cite{ding2018tunable}, and intermediate values of SRO are linearly interpolated. The SRO is sampled using the OTIS code in Ref.~\cite{fey2022random}.}
\label{fig:diffusion}
\end{figure*}

The TKS adopts the reward function of
\begin{equation}
    r_t = -E_b^{\rm NEB} + k_{\rm B}T\log \nu_a^{\rm NEB} ,\ \ \ \nu_a^{\rm NEB} = \frac{\prod_{i=1}^{3M}\nu_i}{\prod_{j=1}^{3M-1}\nu_j^*},
\end{equation}
where $\nu_i$ and $\nu_j^*$ are the $i$th normal mode vibration frequency at state $s_t$ and the $j$th positive vibration frequency at the transition saddle point between $s_t$ and $s_{t+1}$. The model is trained as a contextual bandit problem~\cite{bouneffouf2020survey}, where the value function $Q_\theta (s_t,a_t)$ is trained to fit the instantaneous reward $r_t$ (minimizing $\langle (Q_\theta (s_t,a_t)-r_t)^2\rangle$). Then, as $\Gamma_{s_ta}=\nu_a^{\rm NEB}e^{-E_b^{\rm NEB}/k_{\rm B}T}=e^{r_t/k_{\rm B}T}$ (according to HTST) gives an estimation of the rate of the transition corresponding to action $a$, the policy in Eq.~\eqref{eq:policy} gives the physical transition probability $P(a|s_t)=\Gamma_{s_ta}/\sum_{a'\in \mathcal A_{s_t}}\Gamma_{s_ta'}$. The average residence time of the system on the state $s_t$, $\langle \Delta t\rangle = (\sum_{a\in \mathcal A_{s_t}}\Gamma_{s_ta})^{-1}$, is then estimated as $(\sum_{a\in \mathcal A_{s_t}}e^{Q_\theta (s_t,a)})^{-1}$. Expressing the reward $r_t = r_t^0+r_t^1T$ as a linear function of $T$, the constant term $r_t^0$ and linear term $r_t^1$ can be fitted simultaneously by a two-component value function $(Q_\theta^0,Q_\theta^1)$ in $Q_\theta=Q_\theta^0+Q_\theta^1T$ to make the model applicable to different temperatures:
\begin{equation}
    \theta \leftarrow \theta - \lambda \nabla_\theta\sum_t\left[(Q_\theta^0(s_t,a_t)-r_t^0)^2+T_{\text{tr}}^2(Q_\theta^1(s_t,a_t)-r_t^1)^2\right],
\end{equation}
where $\lambda$ is the learning rate, and $T_{\text{tr}}$, the training temperature, is a hyperparameter that controls the relative importance of the two terms in the loss function. The two components give neural-network predictions for the energy barrier $E_b^{\text{\rm NN}}\equiv -Q_\theta^0$ and attempt frequency 
$\log{\nu_a^{\text{\rm NN}}} \equiv\frac{Q_\theta^1}{k_{\rm B}}$. 

As a testbed, we first apply our method to hydrogen diffusion in pure FCC Cu and Ni. The model is trained on a $4\times 4\times 4$ cubic supercell with 4 randomly sampled hydrogen sites. The model is then deployed to simulate a single hydrogen diffusion in a $3\times 3\times 3$ cubic supercell for 500 timesteps. This system is simulated at temperatures spanning 250 K to 500 K with an interval of 50 K and repeated 50 times for each temperature. The final displacement $\Delta x_i$, total time $t_i$, and temperature $T_i$ of the $i$th simulation trajectory are recorded. The two parameters $D_0$ and $Q$ in the Arrhenius form of diffusivity $D=D_0e^{-Q/k_{\rm B}T}$ are fitted by the maximum likelihood estimation (MLE):
\begin{equation}
    \max_{D_0, Q} \prod_i \frac{4\pi\Delta x_i^2}{(12\pi D_0 t_i e^{-Q/k_{\rm B}T})^{3/2}}\exp{-\frac{\Delta x_i^2}{12D_0t_i}e^{Q/k_{\rm B}T}}.
\end{equation}
The derived $D_0$ and $Q$ are reasonably consistent with the previous experimental measurement, as shown in Table~\ref{table}. The effective activation energy $Q$ in simulation tends to be slightly smaller than the experimental results. That's probably because a small concentration of trapping sites (defects or impurities) in experiments are not considered in simulation, which slightly increases the average energy barriers.

To test the method's capability to capture compositional complexity, we train the RL model on equiatomic CrCoNi medium entropy alloy. The CrCoNi alloy has recently attracted broad interest because of its outstanding fracture toughness and ductility~\cite{doi:10.1126/science.abp8070}. In the CrCoNi solid solution, each metal atom near the hydrogen can be of different atomic species, giving a complex state space. The predicted $E_b^{\rm NN}$ and $\nu_a^{\rm NN}$ are approximately consistent with the values in the training and testing dataset, as shown in Fig.~\ref{fig:validation}, where the data points are distributed close to the diagonal line in the wide range of observed quantities. The standard deviation errors of the model predictions are close in training and testing datasets, confirming that the training data is not overfitted despite the large volume of model parameters.

The hydrogen self-diffusion in CrCoNi is simulated using the trained model running on one hydrogen in a $4\times 4\times 4$ rhombohedral supercell with short-range ordering obtained from Ref.~\cite{ding2018tunable}. The hydrogen displacement as a function of simulation time is shown in Fig.~\ref{fig:diffusion}a under 300 K using 30 repetitions of $\mu$s long-timescale simulations. An approximate function form of $\langle \Delta x^2\rangle \propto t$ is shown by the blue line, and the diffusivity is estimated as $2.84\times 10^{-14}$ m$^2$/s. Similar simulations are implemented for different temperatures, as shown in Fig.~\ref{fig:diffusion}b. The Arrhenius plot shows a good linear relation. The estimated effective activation energy $Q$ equals $0.43\pm 0.01$ eV, and the pre-exponential factor $D_0$ equals $5\pm 2\times 10^{-7}$ m$^2$/s. To our knowledge, these parameters have not been provided in the literature, so we show these results as predictions of our method.

In CrCoNi, short-range ordering (SRO) has significant influences on various properties of the material ranging from hardness~\cite{zhang2020short} and stacking fault energy~\cite{ding2018tunable} to magnetism~\cite{walsh2021magnetically}. We show that the SRO also has an evident influence on the hydrogen diffusivity in CrCoNi, as shown in Fig.~\ref{fig:diffusion}c. The system with SRO under thermal equilibrium (SRO=1) gives approximately two times the hydrogen diffusivity of the fully random configuration (SRO=0), showing that the SRO enhances hydrogen diffusion. That can be explained by the reduction of Cr-Cr bond concentration by the SRO~\cite{ding2018tunable}, as hydrogen transition energy barriers proximate to the Cr-Cr bond are found higher than the average hydrogen transition energy barriers in our calculations. Our results predict that the hydrogen diffusion behavior can also be tuned by the SRO in multi-principle element alloys.

\begin{figure*}[t]
\centering
\includegraphics[width=\linewidth]{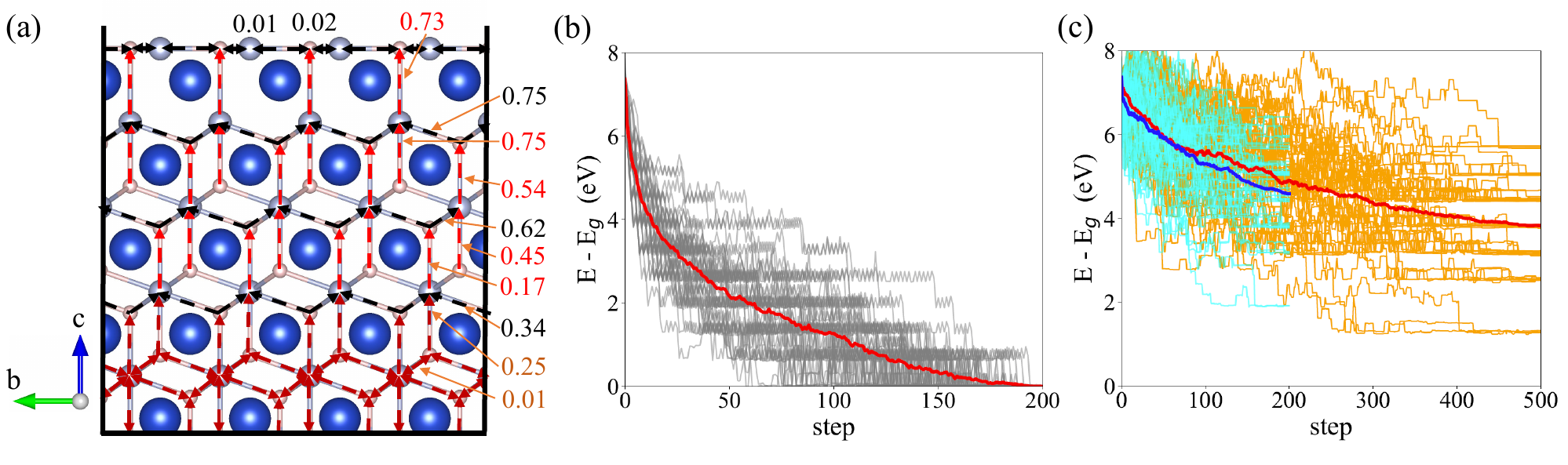}
\caption{Sampling low energy configurations of hydrogen migration to copper (111) surface. (a) Highest probability actions (HPAs) and $Q$ values of hydrogen atoms. The blue, silver, and pink spheres are copper atoms, octahedral interstitial sites, and tetrahedral interstitial sites. The HPAs (the actions with the highest probability according to the policy) are shown by arrows (red arrow: a unique HPA, black arrow: multiple (but not all) actions with equal probabilities, brown arrow: all actions have equal probabilities). The $Q$ values of HPAs are denoted. Energy (using ground state energy as reference) $vs$ simulation step under simulated annealing with (b) $T=1000-950\frac{t}{200} {\rm K}$ using the trained policy and (c) $T=3000-2700\frac{t}{\tau} {\rm K}$ ($\tau$=200  for blue lines and $\tau$=500 for red lines) using Metropolis-Hastings algorithm. The grey/cyan/orange thin lines are 50 simulation trajectories, and the thick red/blue lines are their average.}
\label{fig:anneal}
\end{figure*}

The second working mode of our method, the LSS, sets the reward function as the energy reduction after the transition:
$r_t = E(s_t)-E(s_{t+1})=-\Delta E$. The model is trained by the deep Q network (DQN) algorithm~\cite{mnih2015human}, which aims to maximize the total reward $R=\sum_{t=0}^T \gamma^t r_t$ on a trajectory with a discount factor $\gamma$ close to one (set as 0.8 in our calculation). The model parameters are updated through the Bellman equation~\cite{mnih2015human}:
\begin{equation}
    \theta\leftarrow \theta - \lambda\nabla_\theta\sum_t\left(r_t+\gamma \max _{a^{\prime}} Q_{\theta^t}(s_{t+1},a^{\prime})-Q_\theta (s_t, a_t)\right)^2,
    \label{eq:dqn}
\end{equation}
where $\theta^t$ is the target network that updates less frequently than $\theta$. The converged $Q_\theta (s_t,a_t)$ fits the maximal total rewards after timestep $t$, $\max_{(a_{t+1},a_{t+2},\cdots )} \sum_{t'=t}^T \gamma^{t'-t}r_{t'}$. 
As the $Q$ function ``foresees" the energy reduction of future steps and chooses actions that maximize ``long-term" energy reduction, it is expected to converge to low energy configurations faster than local strategies that only consider single-step energy terms. That provides LSS a simulator of an annealing process, which converges to a near-ground state with fewer timesteps than the TKS.

We demonstrate the LSS's performance in simulating annealing by the hydrogen migration to copper (111) surface process, as shown in Fig.~\ref{fig:anneal}a. $4\times 4\times 3$ hexagonal supercells are constructed with 10 randomly sampled hydrogen sites, and the (111) surface is created with a 15 ${\rm \AA}$ vacuum layer. Hydrogen in the surface adsorption sites has lower energy than that in the bulk interstitial site, so the energy ground state is that all hydrogen atoms are on the surface adsorption sites. However, because of the energy difference between the octahedral sites and tetrahedral sites, the migration pathway involves multiple local energy minimums and low energy barriers, making it challenging to sample the low-energy states~\cite{henkelman2018long}. After training, our RL policy gives the most likely action from each state, as shown in Fig.~\ref{fig:anneal}a. Within the cut-off radius of $8.5 {\rm \AA}$ in Eq.~\eqref{eq:descriptor} from the surface, the highest-probability actions (HPAs) from all sites are oriented towards the surface. The HPAs from surface adsorption sites point to neighbor surface sites. This policy provides orientation for the hydrogen atoms to migrate across the local energy barriers toward the surface sites. The HPAs from sites close to the surface have larger $Q$ values than that far from the surface, as the discount factor reduces the contribution of long-term rewards to the $Q$ function compared to short-term rewards. For sites deeper than the cut-off radius, all move gives the same $Q$ function due to the constraint of symmetry.

We compare the annealing process using the LSS and the Metropolis-Hastings algorithm~\cite{10.1093/biomet/57.1.97}, as shown in Fig.\ref{fig:anneal}b,c. The LSS annealing leads all hydrogen atoms to surface adsorption sites and converges to the energy ground states in 200 timesteps in all 50 trajectories. From the grey lines, one can observe that the system moves across a large number of low-energy barriers and approaches the ground state. In comparison, the Metropolis-Hastings algorithm converges slowly. Less than half of the hydrogen migrates to the surface sites in both 200 and 500 timesteps annealing, leaving $\sim 4$ eV energy above the ground state on average. These results demonstrate that the LSS can show advantageous performance in approaching low-energy configurations compared to straightforward Monte Carlo methods.

\section{Discussion and conclusions}
The TKS and LSS can be viewed as two special cases of a unified DQN framework. The general reward function is:
\begin{equation}
    r_t = -\alpha (\tilde{F}(s^{\text{saddle}}_t) - F(s_t)) - \beta (F(s_{t+1})-F(s_t)),
\end{equation}
where $F(s)\equiv E(s) + k_{\rm B}T\sum_{i=1}^{3M} \log{\nu_i(s)} + F_0$ is the free energy of state $s$, and $\tilde{F}(s^{\text{saddle}})\equiv E(s^{\text{saddle}}) + k_{\rm B}T\sum_{j=1}^{3M-1} \log{\nu_j^*(s^{\text{saddle}})} + F_0$ is the effective free energy of the saddle point ($F_0$ is a state-independent constant). There are three tunable parameters, $\alpha$, $\beta$, and $\gamma$ (in Eq.~\eqref{eq:dqn}), controlling the importance assigned to reproducing the correct transition probability, energy reduction, and long-term performance of the model. The TKS and LSS correspond to $\alpha=1, \beta=\gamma =0$ and $\alpha=0, \beta=1, \gamma \simeq 1$, respectively. Other parameter settings, despite the lack of direct physical interpretation, can be used to explore different configurations in the energy landscape with certain preferences. A probabilistic interpretation of the general framework is discussed in section \RomanNumeralCaps{4}.C, mapping each parameter set to a probability distribution function from which the trajectory is sampled.

Our method provides a computational framework to simulate the long-timescale diffusion and annealing process. Although the simulations in this paper focus on hydrogen diffusion in metals, the method is applicable to diffusion processes in different materials and microstructures, given a specifically designed action space. This method can also bridge large length scales, by first training a model on varied small structures, then deploying the model to guide the long-timescale simulation in a large supercell that includes the complexity of all trained structures. 

\section{Experimental Section}
\subsection{Action space identification algorithm}
The action space $\mathcal A(s) = \{a = (i,\vec{v})\}$ is identified based on the atomic configuration $s$. The algorithm first identifies all hydrogen atoms with indices $i_1, i_2, \cdots$. For each hydrogen atom $i$, the distance of all metal atoms $j$ within a cut-off radius $r_c$ is ranked:
\begin{equation}
    r_{ij_1}\le r_{ij_2}\le \cdots \le r_{ij_M}
\end{equation}
where $r_{ij_k}$ is the distance between atom $i$ and atom $j$. Then, we use all metal atoms $j_k$ with a distance $r_{ij_k}<1.2r_{ij_4}$ (we denote the largest $k$ satisfying the condition as $n$) and the hydrogen atom $i$ itself to construct a convex hull including these atoms. If the hydrogen atom $i$ is a corner of the convex hull, the hydrogen atom is on a surface adsorption site; if the hydrogen atom $i$ is inside the convex hull, the hydrogen atom is a bulk interstitial site.

If the hydrogen atom is in a bulk interstitial site, we choose all face centers, $(\vec{c}_1,\vec{c}_2,\cdots , \vec{c}_m)$, of the convex hull $(j_1,\cdots ,j_n)$. Then, the actions towards every face center $\left(i, \max{(1.6(\vec{c}_k-\vec{r}_i), 1.2{\rm \AA}\frac{\vec{c}_k-\vec{r}_i}{|\vec{c}_k-\vec{r}_i|}})\right), k=1,2,\cdots ,m$ are included into the action space, except there are ``collision" events. The ``collision" event is defined as, if the hydrogen atom $i$ takes the action, it will have a smaller distance than 0.5 ${\rm \AA}$ with at least one other atom. If the hydrogen atom ``collide" with another hydrogen atom, the action is directly discarded. If the hydrogen atom ``collide" with a metal atom, the metal atom will be added to reconstruct a convex hull, and actions towards face centers adjacent to the added atom will be included, except it evokes another ``collision". If that happens, the action will be directly discarded.

If the hydrogen atom is on the surface adsorption site, the convex hull is reconstructed using metal atoms $j_k$ satisfying $r_{ij_k}<1.2r_{ij_3}$. Atoms directly connected with the hydrogen atom, $(j_1, j_2,\cdots ,j_n)$, are identified as the adsorption site (we sort $(j_1, j_2,\cdots ,j_n)$ to form a counterclockwise loop). The adsorption site center is obtained as $\vec{c} = \frac{1}{n}\sum_k \vec{r}_{j_k}$. The adsorption site has $n$ edges, and the $s$th edge center is $\vec{e}_s = (\vec{r}_{j_s}+\vec{r}_{j_{s+1}})/2$. First, the surface diffusion actions $(i,1.6(\vec{e}_s-\vec{c})), s=1,2,\cdots ,n$ are included. Then, the action towards the bulk $\left(i, 3{\rm \AA}\frac{\vec{c}_k-\vec{r}_i}{|\vec{c}_k-\vec{r}_i|}\right)$ is included. If ``collision" happens, the same procedure as the bulk interstitial site case is applied.

\subsection{Detailed parameter settings}
The model training on pure copper and nickel is conducted on $4\times 4\times 4$ cubic supercell of the FCC metals. 3 atomic configurations are generated for each metal, where 4 hydrogen atoms are randomly sampled in all octahedral and tetrahedral sites in each configuration. 20 and 40 trajectories are sampled for copper and nickel, respectively, with 30 timesteps in each. In the atomic relaxation and NEB calculations, all forces converge to $0.05$ ${\rm eV/\AA}$ under the PreFerred Potential (PFP) v4.0.0, which is used throughout this paper. The cut-off radius of the neural network model is 4 ${\rm \AA}$. The embedding network $G_k^1$ has one hidden layer and an output layer both with a size of 12. Throughout the paper, we take the first 1/4 columns of $G_k^1$ to form $G_k^2$, and the input layers of $G_k^{1,2}$ have a size of $N_c+1$, where $N_c$ is the number of element species. We define an element species list: $C = (C_1, C_2, \cdots , C_{N_c}, C_{N_c+1}={\rm action})$, where $C_l$ is the $l$th element. For $G_k^{1,2}(f_c(r_{im}), c_m=C_l)$, the input layer takes the $N_c+1$ dimensional input vector whose $l$th component is $f_c(r_{im})$ and other components are zeros. The fitting network has two hidden layers with a size of 32. The maximum atom number is set as 40, which has not been exceeded during the training. The training temperature is set as 1000 K throughout this paper. After including the $n$th trajectory, one randomly samples a trajectory from probability distribution $P_i = \frac{1-0.99}{1-0.99^{n}}0.99^{n-i}$ (recent trajectory has larger probability) and train 20 gradient descend steps from the sampled trajectory, and repeat this for n times. The training algorithm is Adam throughout this paper, and the learning rate here is set as $10^{-3}$ in all online training. Offline training is conducted to further improve the model's accuracy. We separate the training data into the training dataset (2/3 of the data) and the testing dataset (1/3 of the data). 10000 full gradient descent is implemented on the training dataset. The learning rate changes from $10^{-3}$ to $10^{-5}$ that exponentially decays with timesteps in all offline training in this paper.

The model training on NiCrCo medium entropy alloy is conducted on $4\times 4\times 4$ cubic supercell of the FCC fully random solid solution. 9 atomic configurations are generated for each metal, where 4 hydrogen atoms are randomly sampled in all octahedral and tetrahedral sites in each configuration. 3 independent processes of training are conducted with 101 trajectories in each, and each trajectory contains 30 timesteps. In the atomic relaxation and NEB calculations, all forces converge to $0.05$ and $0.07$ ${\rm eV/\AA}$, respectively. The cut-off radius of the neural network model is 5 ${\rm \AA}$. The embedding network $G_k^1$ has one hidden layer and an output layer both with a size of 24. The fitting network has two hidden layers with a size of 128. The maximum atom number is set as 50, which was not exceeded during the training. The online training parameters are the same as pure metals. As to offline training, we separate the training data the same way as pure metals. Stochastic gradient descent is implemented with a minibatch size of 500 data points (one timestep is a data point). The minibatch is randomly sampled from all data points, and 10 gradient descent steps are applied to each minibatch. That is repeated for 20000 iterations. In order to avoid overfitting, a normalization term of $5\times 10^{-6}\parallel\theta\parallel^2$ is added to the loss function. 

The deep Q learning for copper (111) surface is conducted on $4\times 4\times 3$ hexagonal lattice of FCC copper (4 replications on $a$ and $b$ directions and 3 replications on $c$ direction. $c$ direction is along the 3-fold axis). A vacuum layer of 15 ${\rm \AA}$ is included in the $c$ direction. We implemented 7 independent training processes, 4 of them have only one randomly sampled hydrogen atom in the copper slab (12 configurations are sampled as starting points, and initial configurations are randomly selected from them), and the other 3 have 10 randomly sampled hydrogen atoms (10 configurations are sampled as starting points). 300 trajectories are sampled with 30 timesteps in each. In the atomic relaxation, all forces converge to $0.05$ ${\rm eV/\AA}$. The cut-off radius of the neural network model is 8.5 ${\rm \AA}$, as the model needs more distant atomic information to foresee the long-term rewards. the embedding network $G_k^1$ has one hidden layer and an output layer both with a size of 24. The fitting network has two hidden layers with a size of 128. The maximum atom number is set as 260, which has not been exceeded during the training. After including the $n$th trajectory, one randomly samples a trajectory and trains 5 gradient descent steps from the sampled trajectory, and repeats this for $\lceil n^{2/3}\rceil$ times. The offline training randomly samples a mini-batch with 10 trajectories and applies 10 steps of gradient descent at each iteration. There are 1010 iterations in the training process. 

\subsection{Probabilistic Interpretation of the DQN framework}
By setting the parameters $\alpha$, $\beta$, and $\gamma$, Our method samples different probability distributions. In physical reality, the transition rate is approximately determined by the harmonic transition state theory (HTST):
\begin{equation}
\begin{aligned}
    \Gamma_{s_ta_t} &= \frac{\prod_{i=1}^{3M}\nu_i(s_t)}{\prod_{j=1}^{3M-1} \nu_i^*(s^{\rm saddle}_t)}e^{-(E(s^{\rm saddle}_t))-E(s_t))/k_{\rm B}T} \\
    & = e^{-(\tilde{F}(s^{\rm saddle}_t)-F(s_t))/k_{\rm B}T}
    \label{eq:TST}
\end{aligned}
\end{equation}
At thermal equilibrium, the probability distribution among different states in the state space $\mathcal S$ is:
\begin{equation}
    P(s) = \frac{1}{Z}e^{-F(s)/k_{\rm B}T},  \ \ Z = \sum_{s\in \mathcal S}e^{-F(s)/k_{\rm B}T}
    \label{eq:canonical}
\end{equation}

\subsubsection{$\gamma = 0$: sampling exact transition probabilities}
If $\gamma = 0$, the exact value function $Q^*(s_t,a_t) = r_t = -\alpha (\tilde{F}(s^{\rm saddle}_t)-F(s_t)) - \beta (F(s_{t+1})-F(s_t))$. The problem simplifies into choosing an action based on the next step reward, namely, a contextual bandit problem. If the parameterized $Q_\theta (s,a)$ properly reproduce the exact value function $Q^*(s,a)$, the policy gives:
\begin{equation}
    \pi_\theta (a|s) = \frac{(\Gamma_{sa})^\alpha P(s'_{sa})^\beta}{\sum_{a'\in \mathcal A_s}(\Gamma_{sa'})^\alpha P(s'_{sa'})^\beta}
\end{equation}
where $s'$ is the next state after taking action $a$. For kinetics simulation (TKS) that reproduces the transition probabilities of Eq.~\eqref{eq:TST}, coefficients are set as $\alpha = 1, \beta = 0$. The expected stationary time is then evaluated as:
\begin{equation}
    \tau_t = \frac{1}{\sum_{a}\Gamma_{s_t a}} = \frac{1}{\sum_{a}e^{Q_\theta (s_t a)/k_{\rm B}T}}
\end{equation}
In certain scenarios, the goal is to sample thermal equilibrium distribution. The detailed balance principle proved that the probability distribution follows Eq.~\eqref{eq:canonical} as long as $\alpha +2\beta = 1$. One can set $\beta$ to a larger value to sample more rare transition events while keeping the thermodynamics properties correct. 

\subsubsection{$\gamma \sim 1$: maximizing global probability of a trajectory}
When we set $\gamma \sim 1$, the algorithm maximizes $R(\mathcal T)\simeq \sum_{t=0}^T r_t$ (we consider setting $\gamma$ slightly smaller than 1 as a convergence technique that leads to a small bias). The probability of the trajectory is:
\begin{equation}
    P(\mathcal T) = P(s_0)\prod_{t=0}^{T-1} e^{-\tau_t\Gamma_{s_t}}\Gamma_{s_t\rightarrow s_{t+1}}
\end{equation}
Using the expected value $\tau_t = 1/\Gamma_{s_t}$, the probability becomes $P(\mathcal T|\tau_t = 1/\Gamma_{s_t}) = P(s_0)e^{-T}\prod_{t=0}^{T-1} \Gamma_{s_t\rightarrow s_{t+1}}$. Then, maximizing the total reward corresponds to maximizing:
\begin{equation}
    e^{R(\mathcal T)/k_{\rm B}T-\alpha T}P(s_0)^{\alpha +\beta} = P(\mathcal T|\tau_t = 1/\Gamma_{s_t})^\alpha P(s_T)^\beta
\end{equation}
Here, the initial state $s_0$ does not depend on the policy, so it is constant when doing the maximization. If $\alpha = 0, \beta = 1$, the method aims to sample the most probable final state $s_T$, corresponding to an annealing process that targets the ground state. If $\alpha = 1, \beta = 0$, the method aims to sample the most probable trajectory based on transition kinetics. In the general case, $\alpha$ and $\beta$ can be tuned according to sample probabilities considering both the final state distribution and transition kinetics.

\section{Acknowledgements}
We thank Prof. Cathy Wu, Dr. Shen Shen, and Zhiyuan Shu for their insightful discussions. This work was supported by NSF CMMI-1922206 and DTRA (Award No. HDTRA1-20-2-0002) Interaction of Ionizing Radiation with Matter (IIRM) University Research Alliance (URA). The calculations in this work were performed in part on the Matlantis High-Speed Universal Atomistic Simulator and the Texas Advanced Computing Center (TACC).

\bibliography{bibliography}
%

\end{document}